# Why Superconducting Ta Qubits Have Fewer Tunneling Two-Level Systems at the Air-Oxide Interface Than Nb Qubits

Zhe Wang, Clare C. Yu and Ruqian Wu[*]

*Department of Physics and Astronomy, University of California, Irvine, California 92697, USA*

* Correspondence should be addressed to: wur@uci.edu

Superconducting qubits are a key contender for quantum computing elements, but they often face challenges like noise and decoherence from two-level systems (TLS). Tantalum (Ta) qubits are notable for their long $T_1$ coherence times nearing milliseconds, mainly due to fewer TLS, though the cause was unclear. Our research explored this by analyzing the air-oxide interface with density functional theory, particularly comparing Nb oxide ($Nb_2O_5$) and Ta oxide ($Ta_2O_5$). We discovered that $Ta_2O_5$ forms a smoother surface with fewer dangling O atoms and TLS than $Nb_2O_5$. The greater atomic mass of Ta also lowers the TLS tunnel splittings below the qubit's operating frequency. Furthermore, using external electric fields or $SO_2$ passivation can significantly reduce TLS on Nb surfaces, potentially improving their coherence times.

Superconducting qubits, widely regarded as a leading technology for scalable quantum computing, have seen substantial progress [1,2]. Over the past two decades, their $T_1$ coherence times have dramatically improved, extending from less than 1 ns to hundreds of μs [3,4]. Further development of superconducting qubits with much longer coherence times is crucial for the realization of large-scale fault-tolerant quantum computers. A major obstacle to this enhancement is the presence of two-level systems (TLS) [5-8], typically resulting from the tunneling of an atom [9,10] or a small group of atoms [11] between the minima of a double well potential. TLS with electric dipole moments can couple to both the qubit and microwaves in quantum circuits, leading to dielectric loss [5], charge noise [12], and critical current noise [13]. When the TLS energy splitting matches the qubit energy, the qubit can transfer its excitation energy to the TLS, reducing the qubit's $T_1$ [5,6].

The presence of TLS is predominantly found in the native surface oxide layers of superconducting qubits. While the microscopic nature of these TLS remains elusive, even in conventional qubit materials like aluminum (Al) and niobium (Nb) [4,14-18], TLS do seem to limit the $T_1$ relaxation times of Al qubits to about 70 $\mu s$ [19] and of Nb qubits to about 60 $\mu s$ [15,20]. However, recent studies identified tantalum (Ta) as particularly promising qubit material for quantum computing [21-23]. In 2021 Place *et al.* [21] found that Ta transmon qubits deposited on sapphire had remarkably long coherence times, surpassing 0.3 ms [21] and later reaching an impressive 0.5 ms [22]. The reasons for these enhanced performance metrics in Ta-based devices have not yet been elucidated. As Ta and Nb share similar properties, the difference in their performance in quantum applications is striking. This difference may only be partially attributed to variations in composition at the metal-oxide interface [15,21]. Specifically, while Nb is prone to forming multiple suboxides such as NbO and $NbO_2$ between the Nb metal and the outermost $Nb_2O_5$ layers [14-17,24], Ta consistently forms a highly homogeneous $Ta_2O_5$ layer [22], as also verified by our phase diagram calculations (see Fig. S1 in Supplemental Material [27]).

Recently, the air-oxide interfaces have been identified as more significant sources of TLS [25,26]. This has cast a new light on the puzzling differences between Nb and Ta, despite their outermost oxide layers, $Nb_2O_5$ and $Ta_2O_5$, sharing similar chemical compositions and properties. Studying the differences in TLS formation between Nb and Ta is essential for enhancing quantum computing by optimizing material selection and surface treatments for superconducting qubits. Specifically, exploring their structural and electronic subtleties is crucial, given the high sensitivity of TLS energy levels to local conditions.

In this letter, we explore the potential for TLS to reside on $Nb_2O_5$ and $Ta_2O_5$ surfaces through density functional theory (DFT) calculations (see methods in Supplemental Material [27]). Our findings reveal that the energetically favored $Nb_2O_5$ surfaces are rough with dangling O atoms, which form detrimental TLS with tunnel splittings in the GHz range. Conversely, $Ta_2O_5$ favors a smoother surface with markedly fewer dangling bonds, thereby substantially reducing TLS-related losses. Additionally, the atomic mass of Ta is nearly twice that of Nb, which notably shifts the frequencies of TLS below from the GHz range. These results explain the unexpectedly long coherence times recently observed in Ta-based qubits. [21,22] We also demonstrate that TLS energy splittings can be effectively pushed away from the operating frequencies of qubits by applying a small DC electric field. In addition, the TLS density can be reduced by passivating the surface with sulfur dioxide ($SO_2$) gas. These insights could contribute to further extending the coherence times of superconducting qubits.

Oxides that form on metal surfaces are generally amorphous, making them challenging to model in DFT calculations. Here we use crystalline structures with a reasonably large supercell [9,10,28-30] to mimic amorphous structures. We adopt the most prominent *B*-phase of $Nb_2O_5$ [24,31] and $Ta_2O_5$ [32] (space group *C2/c*, see detailed discussions about its energetic stability in Supplemental Material [27]). As illustrated in Fig. 1(a), the *B*-$M_2O_5$ structure (M = Nb, Ta) features pairs of $MO_6$ octahedra, with adjacent pairs linked through corner-sharing. The optimized lattice parameters for the bulk *B*-$Nb_2O_5$ and *B*-$Ta_2O_5$, as listed in Table S1 of the Supplemental Material [27], deviate by less than 0.9% from the experimental values. We begin by exploring the preferential orientation and termination of the air-oxide interfaces of *B*-$Nb_2O_5$ and *B*-$Ta_2O_5$. We consider their (100), (010), and (001) surfaces with all possible terminations, either stoichiometric (see Fig. S3 in Supplemental Material [27]) or non-stoichiometric. Their surface energies (see Fig. S4 in Supplemental Material [27]), dictate that the most preferable configurations are all stoichiometric, denoted as (100), (010)-III, and (001)-II surfaces, where II and III indicate the position of the terminal O atoms as shown in Fig. 1(a). After exploring possible surface reconstructions (see Fig. S5 in Supplemental Material [27]), we find that the (010)-III surface tends to lower its energy by a 2 × 1 surface reconstruction, which is denoted as the (010)-R surface.

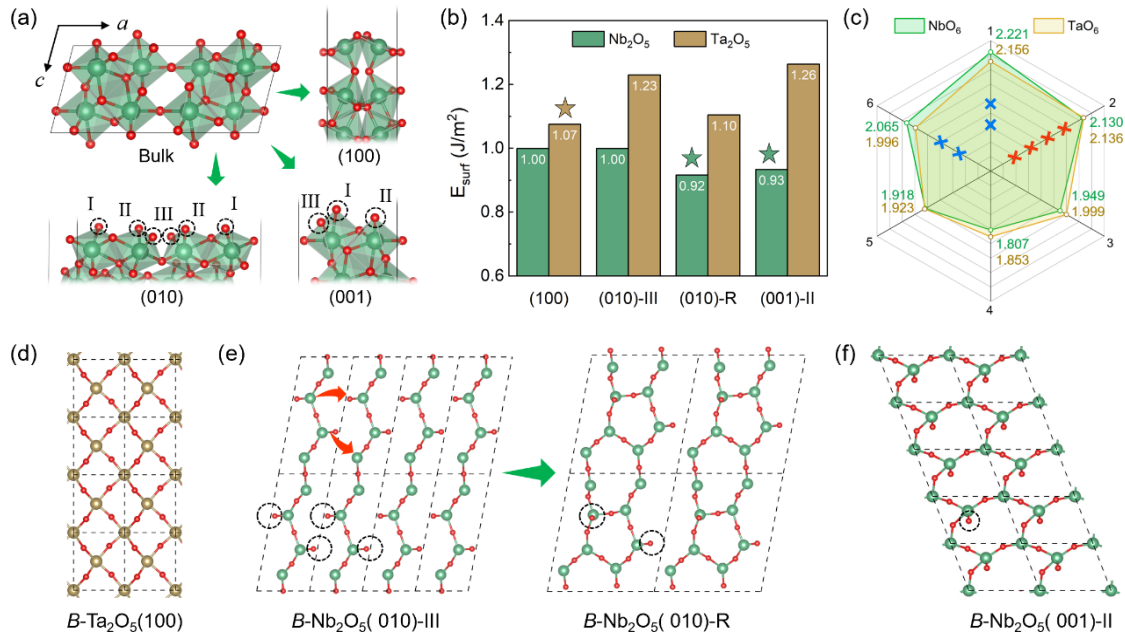

FIG. 1 (color online). *Comparison of surface structures between B-$Nb_2O_5$ and B-$Ta_2O_5$. The Nb, Ta, and O atoms are represented by green, golden, and red balls, respectively. (a) Conventional cell of B-$Nb_2O_5$ and its possible surface structures. The I, II, and III denote different O terminations. (b) Surface energies of the most preferable configurations for each surface orientation, where (010)-R denotes the reconstructed (010) surface. The stars indicate the most stable surfaces, namely the (100) surface for B-$Ta_2O_5$, and the (010)-R and (001)-II surfaces for B-$Nb_2O_5$ which have almost identical surface energies. (c) M-O (M = Nb, Ta) bond lengths of the $MO_6$ octahedron in bulk B-$Nb_2O_5$ (green) and B-$Ta_2O_5$ (gold). The red and blue "×" symbols indicate the bonds that need to be broken when forming the (100) and (001)-II* *surfaces, respectively. (d) (e) (f) Top view of surface morphologies for the most stable surfaces marked in (c). For clarity only the outermost layer is drawn. The black dashed circles denote positions of dangling O atoms.*

The surface energies for all four stoichiometric surfaces of $B$-$Nb_2O_5$ and $B$-$Ta_2O_5$ are given in Figure 1(b). Notably, $B$-$Nb_2O_5$ exhibits a preference for either the reconstructed (010)-R surface or the (001)-II surface, while $B$-$Ta_2O_5$ primarily favors the (100) surface. As shown in Figure 1(d), the $B$-$Ta_2O_5$(100) surface is exceptionally smooth and lacks dangling O atoms. In contrast, the outermost Nb atoms on the $Nb_2O_5$(010)-R surface form alternating five- and seven-membered rings with bridging O atoms, resulting in two dangling O atoms. Additionally, the $Nb_2O_5$(001)-II surface features one dangling O atom on each six-membered ring, as depicted in Figure 1(f). The stoichiometric nature of these surfaces ensures that these dangling O atoms are non-magnetic and stable, as they do not have unpaired electrons.

As $B$-$Nb_2O_5$ and $B$-$Ta_2O_5$ share isomorphic bulk structures and their chemical properties are largely similar, it is intriguing to explore what drives their different surface preferences. To this end, we illustrate the M-O bond lengths within the $MO_6$ octahedrons of bulk $B$-$M_2O_5$ in Figure 1(c). Notably, the Ta-O octahedron appears more isotropic, with a smaller difference between the shortest (b-4) and longest (b-1) bond lengths (0.303 Å) compared to the Nb-O octahedron (0.414 Å). As indicated by the "×" symbol in Fig. 1(c), forming the (100) surface breaks four long b-2 bonds, while forming the (001)-II surface breaks two b-1 and two b-6 bonds. Notably, b-1 and b-6 bonds are significantly shorter (and thus stronger) in $TaO_6$ than in $NbO_6$, whereas the b-2 bonds are nearly identical in both. This subtle difference between the local structures of $B$-$Nb_2O_5$ and $B$-$Ta_2O_5$ is sufficient to explain their distinct surface preferences, as breaking the longer, weaker M-O bonds is energetically more favorable. As the (010)-R surface has dangling O atoms while the (100) surface has none, it is crucial to investigate if dangling O atoms lead to the formation of TLS and hence are responsible for the drastic difference in $T_1$ coherence times that are observed between Nb and Ta superconducting qubits.

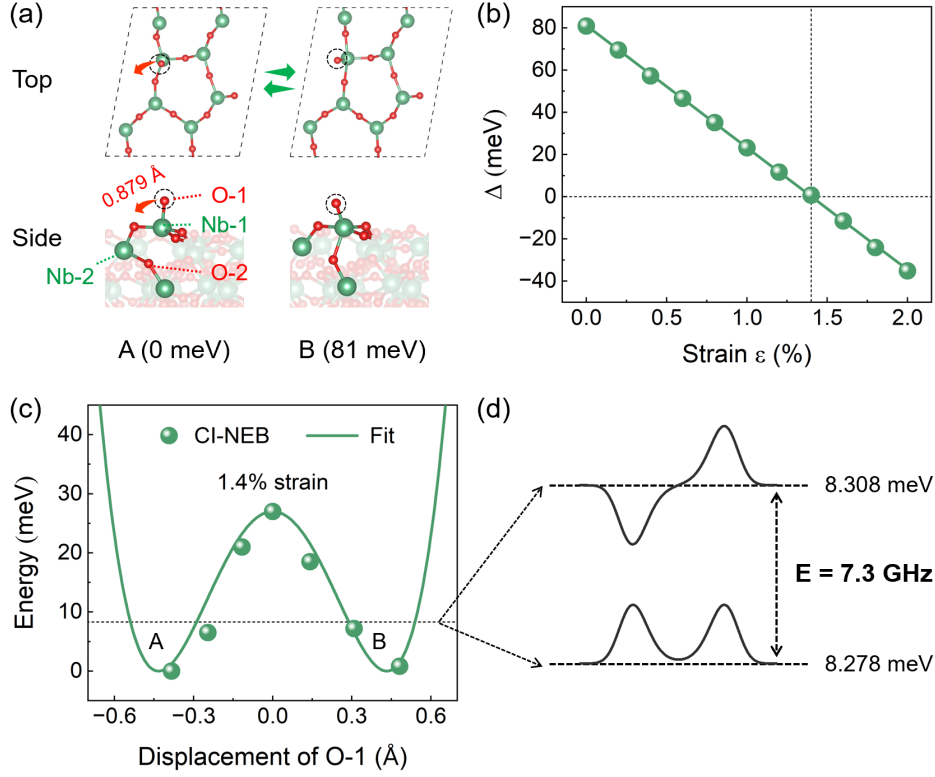

FIG. 2 (color online). A typical TLS formed on the (010)-R surface of B-Nb$_2$O$_5$. (a) Top and side views of the two local stable states A and B. In the top view, only the outermost layer is drawn for clarity. The black dashed circles denote positions of the rotated dangling O atom. (b) Energy difference $\Delta$ between the two local stable states as a function of the biaxial strain $\varepsilon$. (c) Double well potential formed by structure A and B under a 1.4% strain. The green dots are the DFT energies calculated using the CI-NEB method, and the green line indicates the fitted curve using Eq. (1). (d) The tunneling splitting of the TLS, estimated by numerically solving the one-dimensional Schrödinger equation using the symmetric double well potential in (c). The wavefunctions are indicated by black lines.

After carefully moving the dangling O atoms and their neighbors in the supercell from the optimized geometries (The details of this process are in the Supplemental Material [27]), we successfully identify two TLS on the *B*-Nb$_2$O$_5$(010)-R surface within the limit of the unit cell size. The first TLS (TLS-I) primarily involves the rotational motion of the dangling O atom around the perpendicular axis passing through the adjacent Nb-1 atom. As illustrated in Fig. 2(a), we find a double well potential between two stable states, labelled as A and B, with an energy asymmetry of Δ=81 meV at zero strain. It is important to note that the rotation of O-1 alone is insufficient to establish the metastable state B. The transition from state A to B is characterized by a joint motion of two O atoms, a shift of 0.879 Å for the dangling O-1 atom and a shift of 0.485 Å for the underneath O-2 atoms. The joint process of bond breaking between O-2 and Nb-2 followed by formation of new bond between O-2 and Nb-1 is essential for stabilizing structure B.

The TLS energy splitting can be estimated from Δ along with tunneling matrix element $\Delta_0$ as $E_{\text{TLS}} = \sqrt{\Delta^2 + \Delta_0^2}$. The value of Δ=81 meV appears to be too large to be in resonance with superconducting qubits operating in the GHz region. [6-8] However, the amorphous nature in real surface oxides introduce random local strains, which are expected to significantly alter Δ [33], giving rise to a broad distribution of $E_{\text{TLS}}$ [7,8]. Indeed, we find that the value of Δ is very sensitive to a biaxial strain ε. As shown in Fig. 2(b), the TLS-strain coupling coefficient, $\gamma = \frac{1}{2}\frac{\partial \Delta}{\partial \varepsilon}$, is as large as 2.89 eV. This value is larger than that measured in Al superconducting qubits (0.1 - 1 eV) [34] but is close to that estimated for $SiO_2$ [35]. As a result, the double well potential can become symmetric as shown in Fig. 2(c) with a local stretching strain of 1.4%. Using the climbing image-nudged elastic band (CI-NEB) method [36], we calculate the double well potential for the transition path from A to B and fit it with a quartic polynomial

$$U(x) = V - \frac{8V}{d^2}x^2 + \frac{16V}{d^4}x^4, \qquad (1)$$

where $V$ is the height of the energy barrier and $x$ is the position of the dangling O-1 atom. From our fits, we can determine the distance between the two minima $d$ (= 0.865 Å) as well as the barrier $V$ (= 27 meV).

Now, we can estimate the minimal tunnel splitting of TLS-I by solving the one-dimensional (1D) Schrödinger equation for the symmetric double well potential,

$$-\frac{\hbar^2}{2m}\frac{d^2\psi}{dx^2} + U(x)\psi(x) = E\psi(x) \qquad (2)$$

Here, we use the mass of the dangling O atom, $m$ = 16 amu, to find the eigenvalues and wavefunctions of the two tunneling states. As shown in Fig. 2(d), the tunnel splitting is 7.3 GHz, which is in the experimental range. The motion of dangling O atom, with a negative charge of -0.8 e from Bader charge analysis, leads to a significant change in the electric dipole moment of 5.33 Debye between the TLS states A and B. This is consistent with experimental observations, which are typically distributed around 2.8 and 8.3 Debye.[33,37].

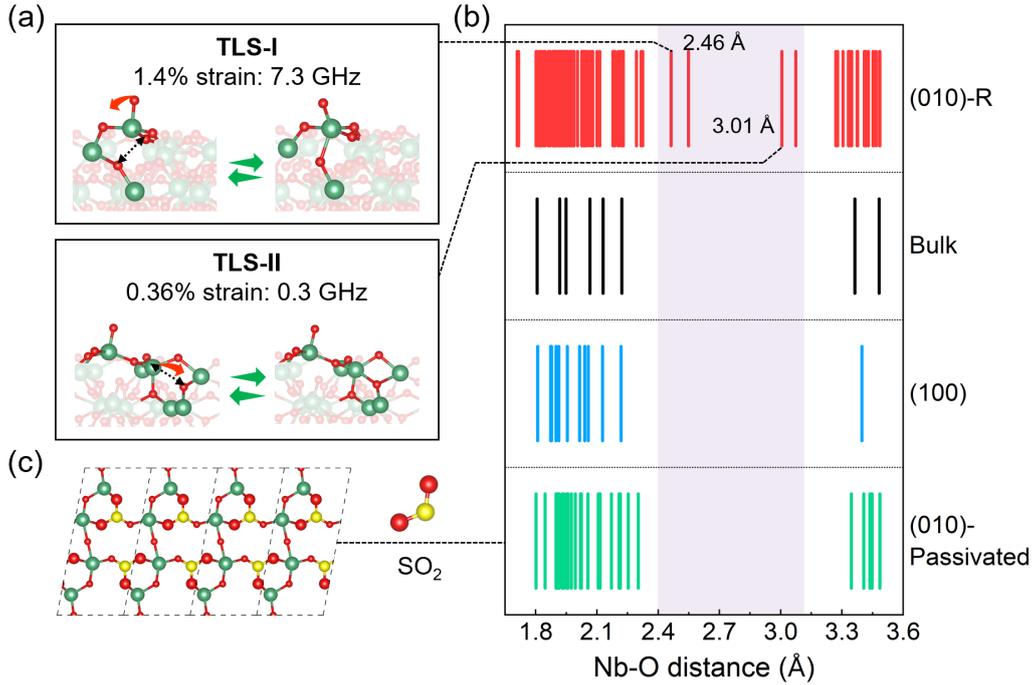

FIG. 3 (color online). Structure of two TLS on *B*-Nb$_2$O$_5$ with a proposed way to reduce their density. (a) Two typical TLS on the (010)-R surface of *B*-Nb$_2$O$_5$. The local strains that lead to $\Delta = 0$ are explicitly labeled, along with the corresponding tunnel splitting frequencies of the TLS. The dotted arrows indicate the intermediate bonds marked in (b). (b) Distribution of the Nb-O bond lengths in the (010)-R surface, bulk, (100) surfaces, and SO$_2$ passivated (010) surface of *B*-Nb$_2$O$_5$. The light purple area denotes the intermediate bonds, defined as bonds with lengths in the range of 2.4 to 3.1 Å. (c) Top view of the SO$_2$ passivated (010) surface, where the S and O atoms of SO$_2$ are represented by yellow and the larger red balls, respectively.

Another TLS on the *B*-Nb$_2$O$_5$(010)-R surface is TLS-II (see Fig. 3(a)), which involves hopping of a Nb atom close to the dangling O atom even though this O atom seldom moves. A local strain of 0.36% produces a symmetric double well potential with a barrier width $d = 0.485$ Å and an energy barrier height $V = 28$ meV. As the motion of Nb plays the pivotal rule in the formation of TLS-II, we use the atomic mass of Nb atoms ($m = 92.9$ amu) for solving the 1D Schrödinger equation for the symmetric double well potential. This gives a tunnel splitting of 0.3 GHz and an electric dipole moment change of 5.00 Debye for TLS-II.

We may further estimate the TLS density on the $Nb_2O_5$ surface. The standard TLS model assumes that $\Delta$ and $V$ are uniformly distributed [7,8]. Equivalently, we also assume a flat distribution of local strains $\varepsilon$ in the range of 0%~5%. Consider an extreme case with all the surface dangling O atoms forming TLS on the (010)-R surface, where the TLS are separated by an average distance of $a = 12$ Å. We obtain an upper bound to the TLS density per GHz and per unit area as

$$\sigma = \sum_i \frac{S/a^2}{S(0.05 \times 2\gamma_i)} = \sum_i \frac{1}{0.1\gamma_i a^2}, \tag{3}$$

where $S$ is the surface area, and $i$ runs over TLS-I and TLS-II. Using the calculated strain coupling coefficient $\gamma_{TLS-I} = 2.89$ eV and $\gamma_{TLS-II} = 2.91$ eV, we arrive at $\sigma \approx 20/(GHz \cdot \mu m^2)$, which is an order of magnitude larger than the experimental value of 0.4~4.4 /(GHz·μm$^2$) observed in Al/AlO$_x$/Al and Nb/AlO$_x$/Nb junctions [7]. Assuming a surface layer thickness of 3 nm, 20/(GHz·μm$^2$) corresponds to a bulk density of $1 \times 10^{46}$/(J·m$^3$), a value consistent with TLS densities found in typical glasses [38].

Both types of TLS involve breaking Nb-O bonds, which generally involves large energies. This seems paradoxical considering the relatively low energy barriers (in the tens of meV) of their double well potentials. To understand this, Fig. 3(b) illustrates the distribution of Nb-O bond lengths across various surfaces: (010)-R, (100), and within bulk $B$-$Nb_2O_5$. As highlighted by the light pink area in the figure, there is a significant gap between the first and second neighbor bond length groups in bulk $B$-$Nb_2O_5$ and on the smooth (100) surface, indicating their strong structural stability. However, for the (010)-R surface that features dangling O atoms, several intermediate bond lengths fall into the aforementioned gap, as indicated by dashed lines in Fig. 3. These lengths range from 2.4 to 3.1 Å, suggesting that these O (Nb) atoms are only loosely bound and may jump between different Nb (O) atoms. The presence of intermediate bond lengths indicates weak Nb-O bonds on these surfaces, which can be broken or reformed with relatively low energies, thus accounting for the small energy barriers of the TLS. The emergence of these intermediate bonds is closely linked to the presence of dangling O atoms, which are attached to only one Nb atom and thereby extract more electrons from it than typically seen in the bulk structure. Consequently, the Nb atoms beneath the surface cannot supply a sufficient number of electrons to the subsurface O atoms, resulting in overall weaker Nb-O bonds in the surface region.

The observation that the surface energy of $Ta_2O_5$ (010)-R is also close to that of $Ta_2O_5$ (100) in Fig. 1(b) might suggest the presence of dangling O atoms and TLS on Ta-based qubits as well. However, DFT calculations reveal notable differences in the TLS configurations on $Ta_2O_5$(010)-R compared to $Nb_2O_5$(010)-R, particularly in the heights and widths of their potential wells. For instance, the barrier height of TLS-II becomes 39 meV on $Ta_2O_5$(010)-R, much higher than that on $Nb_2O_5$(010)-R which has $V = 28$ meV (see Fig. S6 and Table S2 of the Supplemental Material [27]). Furthermore, the heavier atomic mass of Ta ($m = 180.9$ amu) compared to Nb significantly reduces the tunneling probability and TLS energy splitting, according to the Wentzel-Kramers-Brillouin (WKB) theory [39], which gives $E_{TLS} \propto e^{-d\sqrt{2mV/\hbar^2}}$. Using parameters derived from DFT calculations, we find that the tunnel splitting of TLS-II on the $B$-$Ta_2O_5$ (010)-R surface is only 66 kHz. This value is significantly lower than those typically observed in the GHz region, which is relevant for the operating frequencies of quantum computing devices.

We can now explain why Ta-based qubits may significantly outperform those based on Nb. The primary reason is that $Ta_2O_5$ features a highly stable and smooth (100) surface, which hosts far fewer defect-associated TLS. This characteristic is crucial because TLS are known sources of quantum decoherence and noise in qubits. The second key factor contributing to the superior performance of Ta-based qubits involves several intertwined physical properties: 1) heavy atomic mass; 2) large spatial separation between metastable structures; and 3) high energy barrier. These properties significantly lower the tunneling probability and drive the tunneling splitting of probable TLS away from the frequency range critical to qubit operations. Consequently, the operational efficiency of Ta-based qubits is substantially enhanced, making them more promising for quantum computing applications.

Finally, the influence of TLS on the surfaces of oxides can be mitigated by employing DC electric fields. This approach leverages the electric dipole moments associated with TLS. As demonstrated in Fig. S7(a), the asymmetry energies of two distinct types of TLS vary linearly with the applied electric field, which was also reported in references [33,40]. Using the field dependent potential wells in Eq. (2), we find that $E_{TLS}$ can be effectively driven out of the GHz range, as depicted in Fig. S7(b). Additionally, we also calculate the resonant dielectric absorption coefficient [41] which is proportional to the square of the dipole moment (see Fig. S7(b)). A very recent experiment [42] reported a significant 23% improvement in $T_1$ of a superconducting qubit by applying DC electric fields. However, it is important to note that while applying an electric field can shift certain TLS out of resonance, it may inadvertently bring other, previously non-resonant TLS into the GHz range. [33]

Another potential method to reduce the density of TLS on oxide surfaces involves eliminating dangling O atoms through chemisorption. After exploring various gases for this purpose, we find that the adsorption of sulfur dioxide ($SO_2$) is particularly effective, especially on the (010)-R surface of Nb. $SO_2$ chemisorption can effectively "cap" the dangling O atoms, thereby stabilizing the surface and reducing sites where TLS can form. As depicted in Fig. 3(c), $SO_2$ molecule may bridge the dangling O atom and its two Nb neighbors, restricting their motion to form metastable configurations. The adsorption energy of $SO_2$ on $Nb_2O_5$(010)-R is high, 1.58 eV per molecule. This stable chemical intervention results in the elimination of intermediate bond lengths, as illustrated in the bottom panel of Fig. 3(b). As a result, all atoms are effectively frozen in the ground state geometry and the TLS density is hence significantly reduced.

In summary, our study offers a novel perspective on the disparities in $T_1$ relaxation times between Nb and Ta superconducting qubits, with a particular focus on the presence of TLS at their air-oxide interfaces. Our findings reveal that $Nb_2O_5$ tends to prefer rougher surfaces which host dangling O atoms, contributing to the formation of TLS and thus affecting the coherence times of Nb qubits. In contrast, $Ta_2O_5$ maintains a smoother (100) surface that closely resembles a bulk-like environment where dangling O atoms are rare, resulting in fewer TLS-related losses and, consequently, longer coherence times in Ta qubits. Furthermore, the heavier atomic mass of Ta also contributes by lowering the tunnel splitting and driving TLS away from the GHz frequency range of qubits. These insights emphasize the critical role that surface conditions—specifically, the presence or absence of dangling bonds—play in determining the performance of superconducting qubits. Additionally, we explore promising strategies for mitigating the deleterious effects of TLS. Applying a DC electric field and passivating the surface with $SO_2$ gas have both shown potential in altering or eliminating TLS on qubit surfaces. These approaches could significantly advance the development of superconducting qubits by enhancing their coherence times and overall operational stability, paving the way for more reliable and effective quantum computing technologies.

This work was supported by the USA-DOE, Office of Basic Energy Science (Grant No. DE-FG02-05ER46237). It was also supported in part by LPS/ARO grant W911NF24C0001. Calculations were conducted on the High-Performance Community Computing Cluster at UC Irvine and were also partially performed on supercomputers at The National Energy Research Scientific Computing Center.

# Supplementary Material for "Why Superconducting Ta Qubits Have Fewer Tunneling Two-Level Systems at the Air-Oxide Interface Than Nb Qubits"

Zhe Wang, Clare C. Yu, and Ruqian Wu[*]

*Department of Physics and Astronomy, University of California, Irvine, California 92697, USA*

## Computational Methods

Our density functional theory (DFT) calculations were performed using the projector augmented wave (PAW) [1] method as implemented in the Vienna *ab* initio simulation package (VASP) [2,3]. The exchange-correlation functional was described by the generalized gradient approximation (GGA), in the form proposed by Perdew, Burke, and Ernzerhof (PBE) [4,5]. An energy cutoff of 550 eV was used for the plane-wave basis expansion. In the bulk calculations, a Γ-centered Monkhorst-Pack **k**-point mesh of $2 \times 5 \times 4$ was adopted for the Brillouin-zone integration, and both the lattice constants and atomic positions were fully relaxed until the force acting on each atom was less than 0.01 eV/Å. While in the subsequent slab calculations, the in-plane lattice constants were fixed to their bulk values. When exploring the two-level system (TLS) formed by dangling O atoms, the climbing image nudged elastic band (CI-NEB) [6] method was used to determine the barrier height of the double-well potential. It has been reported that the PBE functional without van der Waals (vdW) corrections predicts a wrong ground state phase for $Nb_2O_5$ and $Ta_2O_5$. We therefore included the vdW correction using the DFT-D3 [7] method with the Becke-Johnson damping function [8] in all of the calculations. Structure visualization was performed using the VESTA [9] package.

## Phase Diagram of Nb and Ta Oxides

Experiments find that the native oxide layer of Nb has a complex composition. As schematically depicted in in Fig. S1(a), it is composed of multiple suboxides (NbO and NbO$_2$) close to the Nb side, followed by Nb$_2$O$_5$ as the outermost layer [10-14]. By comparison, the native oxide layer of Ta consists solely of Ta$_2$O$_5$ [15] (Fig. S1(b)).

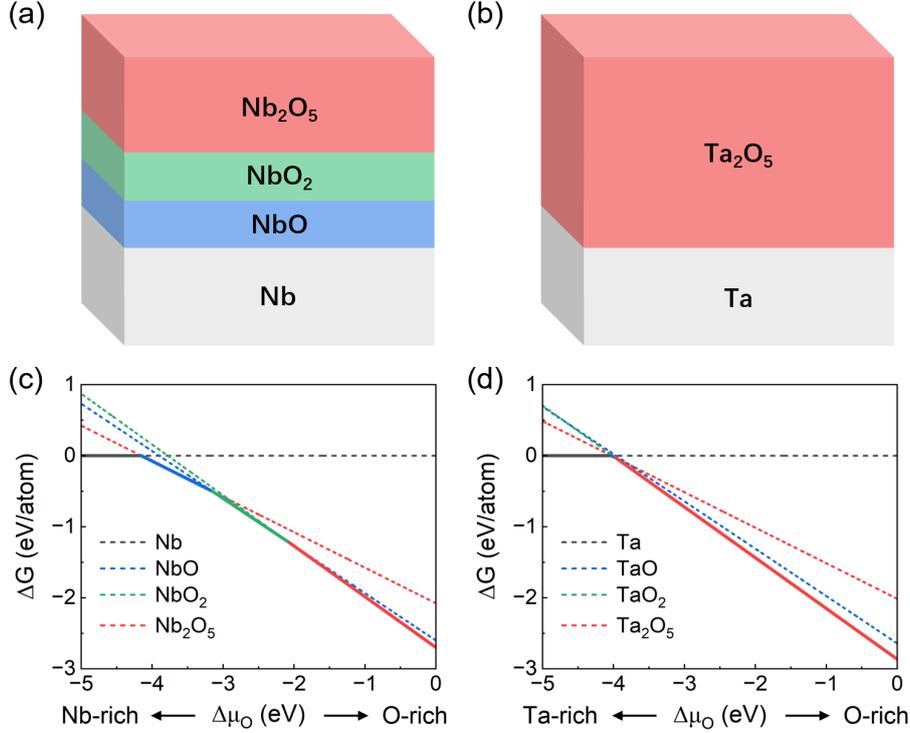

FIG. S1. (a) (b) Schematic diagram of the native oxide layers of Nb and Ta. The Nb oxides are composed of multiple components, including NbO, NbO$_2$, and Nb$_2$O$_5$; while Ta oxides consist solely of Ta$_2$O$_5$. (c) (d) Phase diagram of Nb and Ta oxides, obtained by comparing the Gibbs free energy of formation (ΔG) under different chemical potential of O elements (Δμ$_O$).

This is also confirmed by our DFT phase diagram calculations. We calculate the Gibbs free energy of formation (ΔG) of different Nb and Ta oxides using

$$\Delta G = \frac{1}{x+y}\left[E_{\mathrm{M}_x\mathrm{O}_y} - xE_\mathrm{M} - y\left(\frac{1}{2}E_{\mathrm{O}_2} + \Delta\mu_\mathrm{O}\right)\right]. \tag{S1}$$

Here, M represents Nb or Ta elements; $E_{\mathrm{M}_x\mathrm{O}_y}$, $E_\mathrm{M}$, and $E_{\mathrm{O}_2}$ are the calculated total energy of bulk M oxides, bulk M metal, and O$_2$ molecule, respectively; Δμ$_O$ is the chemical potential of O with respect to half the total energy of an O$_2$ molecule. For each M oxide, we consider various polymorphic forms and select the most stable one, i.e., MO, MO$_2$, and M$_2$O$_5$ with space group $Pm\bar{3}m$, $I4_1/a$, and $C2/c$, respectively. As shown in Fig. S1(c), when Δμ$_O$ changes from the O-rich to Nb-rich conditions, the most stable phase of Nb oxides varies from Nb$_2$O$_5$ to NbO$_2$, and then to NbO. In contrast, the most stable phase of Ta oxides is always Ta$_2$O$_5$, regardless of the O chemical potentials.

## Selection of the Representative Structures for Nb$_2$O$_5$ and Ta$_2$O$_5$

As discussed in the main text, we adopt the crystalline structure of Nb$_2$O$_5$ and Ta$_2$O$_5$ to investigate the oxides of Nb and Ta at the air-oxide interface, since the real amorphous structures are difficult to be modelled. The crystalline structures of Nb$_2$O$_5$ and Ta$_2$O$_5$, however, exhibit a number of polymorphs. The one we selected should have a high energetic stability, as well as a simple structure to facilitate our theoretical simulations. Based on these considerations, we selected one specific polymorph as the representative model for each metal oxide.

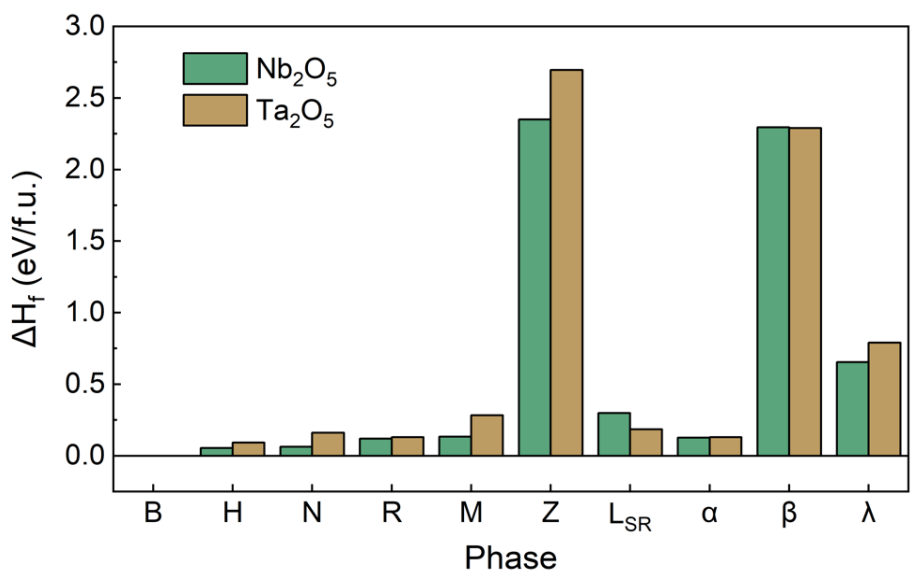

FIG. S2. Formation energies per formula unit (f.u.) of various Nb$_2$O$_5$ and Ta$_2$O$_5$ polymorphs with respect to their *B*-phases. The zero-point energy and finite temperature effects are not taken into account.

Nb$_2$O$_5$ is known to crystallize in various polymorphic forms [16], such as the *B* (or *ζ*), *H*, *N*, *R*, *M*, and *Z* phases. Their relative stabilities are compared in Fig. S2 using the DFT calculated formation energies. Although *H*-Nb$_2$O$_5$ is believed to be the most thermodynamically stable phase under high temperatures [17,18], our calculation indicates that *B*-Nb$_2$O$_5$ is the most stable one under low temperatures, in accordance with previous DFT studies [18]. Considering also that the primitive cell of *H*-Nb$_2$O$_5$ contains 98 atoms, while *B*-Nb$_2$O$_5$ has only 14 (28) atoms in its primitive (conventional) cell which is more suitable for DFT calculations, we adopt *B*-Nb$_2$O$_5$ to investigate the surface oxides of Nb films. The calculated lattice parameters of *B*-Nb$_2$O$_5$ are listed in Table S1, less than 0.7% deviation from the experimental values.

For Ta$_2$O$_5$, our calculation indicates that the *B*-Ta$_2$O$_5$ is the most stable phase (Fig. S1), in consistent with recent DFT studies [19]. Hence, we utilize *B*-Ta$_2$O$_5$ to mimic surface oxide on Ta films. It is important to note that we have the same polymorphic form for Nb$_2$O$_5$ and Ta$_2$O$_5$, which allows us to reveal the difference between Nb and Ta more intrinsically without being concerned with the structural difference. The calculated lattice parameters of *B*-Ta$_2$O$_5$ are listed in Table S1, less than 0.9% deviation from the experimental values.

TABLE S1. Calculated lattice parameters of *B*-Nb$_2$O$_5$ and *B*-Ta$_2$O$_5$. The experimental values are also listed for comparison.

| Oxide | | $a$ (Å) | $b$ (Å) | $c$ (Å) | $\alpha$ | $\beta$ | $\gamma$ |
|---|---|---|---|---|---|---|---|
| *B*-Nb$_2$O$_5$ | Calc. | 12.82 | 4.89 | 5.57 | 90° | 104.7° | 90° |
| | Expt. [20] | 12.74 | 4.88 | 5.56 | 90° | 105.0° | 90° |
| *B*-Ta$_2$O$_5$ | Calc. | 12.89 | 4.89 | 5.55 | 90° | 103.8° | 90° |
| | Expt. [21] | 12.79 | 4.85 | 5.53 | 90° | 104.3° | 90° |

**Calculation of Surface Energies**

The surface structures of the metal oxides are simulated by slab models. We start from fully relaxed bulk structures and cleaved them into slabs in such a way that the surfaces at each side are equivalent. The slab thicknesses are chosen to be more than 10 Å, thick enough to avoid interactions between the two sides. A vacuum space of 15 Å along the *z* axis is inserted to eliminate interactions between the periodically repeated images. The surface energy is calculated by

$$E_{\text{surf}} = \frac{1}{2A}(E_{\text{slab}} - n_M \mu_M - n_O \mu_O), \quad (S2)$$

where $E_{\text{slab}}$ is the total energy of the slab; $\mu_M$ and $\mu_O$ are the chemical potentials of M atoms (M = Nb, Ta) and O atoms, respectively; $n_M$ and $n_O$ are their corresponding atom numbers; and $A$ is the surface area. The values of $\mu_M$ and $\mu_O$ are dependent on experimental conditions. At the air-oxide interface, the $O_2$ gas in the atmosphere acts as a reservoir of O elements in the metal oxides, and thus the chemical potential of O with respect to $1/2 E_{O_2}$ can be expressed as

$$\Delta \mu_O(T,P) = \mu_O(T,P) - \frac{1}{2}E_{O_2} = \frac{1}{2}[H(T,P) - TS(T,P)], \quad (S3)$$

where $E(O_2)$ is the DFT calculated total energy of $O_2$ molecule; $H$ and $S$ are the enthalpy and entropy of $O_2$ gas phase at temperature $T$ and pressure $P$. Given that the oxide layers are formed under ambient conditions, we can simply use the tabulated values [22] for $O_2$ at standard states ($H_0$ = 8.68 kJ·mol$^{-1}$, $S_0$ = 205.138 J·mol$^{-1}$·K$^{-1}$ at $T_0$ = 298.15 K, $P_0$ = 1 bar), which yields $\Delta \mu_O$ = -0.272 eV. The chemical potential of the metal element $\mu_M$ can then be determined via the thermodynamic stability conditions

$$\begin{aligned} E_{\text{Nb}_2\text{O}_5} &= 2\mu_{\text{Nb}} + 5\mu_O, \\ E_{\text{Ta}_2\text{O}_5} &= 2\mu_{\text{Ta}} + 5\mu_O. \end{aligned} \quad (S4)$$

Using Eqs. (1) and (3), one can find that for stoichiometric surfaces, the surface energy is a constant regardless of the chemical potentials, as it can be written as

$$E_{\text{surf}} = \frac{1}{2A}(E_{\text{slab}} - nE_{\text{bulk}}), \quad (S5)$$

where $E_{\text{bulk}}$ is the energy of bulk phase per unit cell, and $n$ is the number of unit cells that the slab contains.

### Surface Structures of *B*-Nb$_2$O$_5$ and *B*-Ta$_2$O$_5$

We consider the (100), (010), and (001) surfaces for *B*-Nb$_2$O$_5$ and *B*-Ta$_2$O$_5$, and take into account all possible terminations, either stoichiometric or non-stoichiometric. It is worth emphasizing that *B*-Nb$_2$O$_5$ has several important "hidden" surface configurations, which were all neglected in previous DFT studies [23]. Here we take its stoichiometric surface for example. As displayed in Fig. S3, the (100) orientation has the simplest surface with only one stoichiometric configuration. By comparison, the (010) orientation shows much more complex surface structures. As displayed in Fig. S3(b), disconnecting the NbO$_6$ octahedra along the (010) direction leaves six O atoms on its surface, which are located at the high (I), medium (II), and low positions (III). In order to form a stoichiometric surface, only two O atoms should be retained, leading to three surface configurations, namely (010)-I, -II, and -III. Likewise, the (001) orientation has three stoichiometric surfaces as shown in Fig. S3(c), where the outermost O atom can be stabilized at three different positions, denoted as (001)-I, -II, and -III.

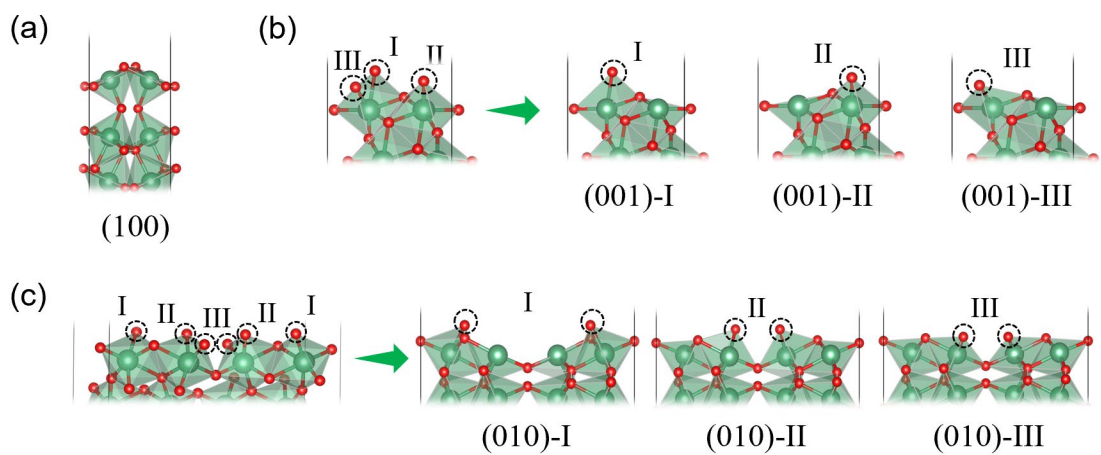

FIG. S3. Structures of the seven possible stoichiometric surfaces of *B*-Nb$_2$O$_5$. The Nb and O atoms are represented by green and red balls, respectively. The black dashed circles indicate the position of dangling O atoms.

In addition to the above seven stoichiometric surfaces, we also construct another 26 possible non-stoichiometric surfaces for *B*-Nb$_2$O$_5$ and *B*-Ta$_2$O$_5$. The surface energies of all these 33 surfaces under ambient conditions are calculated using Eqs. (S2)-(S5), and the results are shown in Fig. S4(a). We find that the most preferable configurations for each surface orientation are all stoichiometric for both *B*-Nb$_2$O$_5$ and *B*-Ta$_2$O$_5$.

Specifically, they are the (100), (010)-III, and (001)-II surfaces as displayed in Fig. S3. Note that the (001)-II surface was overlooked in previous DFT studies [23], presumably because its construction is not straightforward as mentioned before.

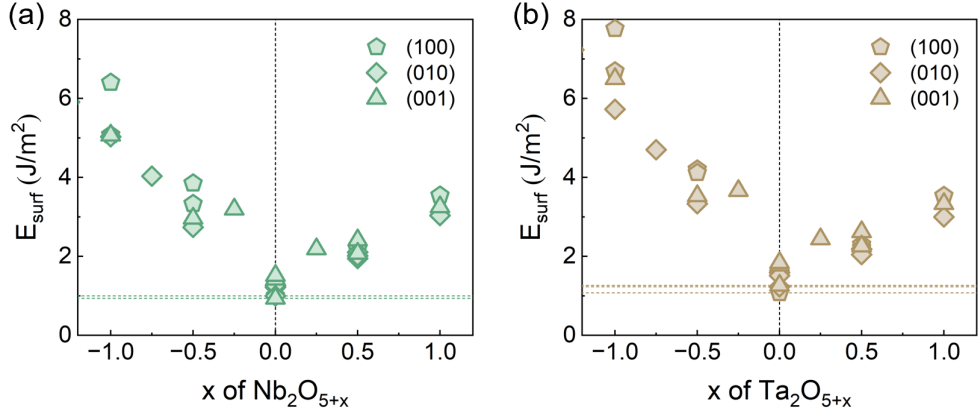

FIG. S4. Surface energies of all possible surfaces for (a) *B*-$Nb_2O_5$ and (b) *B*-$Ta_2O_5$. Different polygons represent different surface orientations. The horizontal dashed lines indicate the most stable configuration of each surface orientation.

We then examine whether the three surfaces will further lower their energies by surface reconstruction. For the (100) and (001)-II surfaces, we manually construct varies reconstructed structure, but they spontaneously relaxed back, indicating the reconstruction is unfavored. While for the (010)-III surface, we identify various reconstructed structures as show in Fig. S3, where the outermost Nb atoms form different rings through the bridging of O atoms. The one with the lowest energy is formed by alternative five- and seven-membered rings, which lower the surface energy of the (010)-III surface by 0.083 J/m$^2$ and 0.125 J/m$^2$ for *B*-$Nb_2O_5$ and *B*-$Ta_2O_5$, respectively.

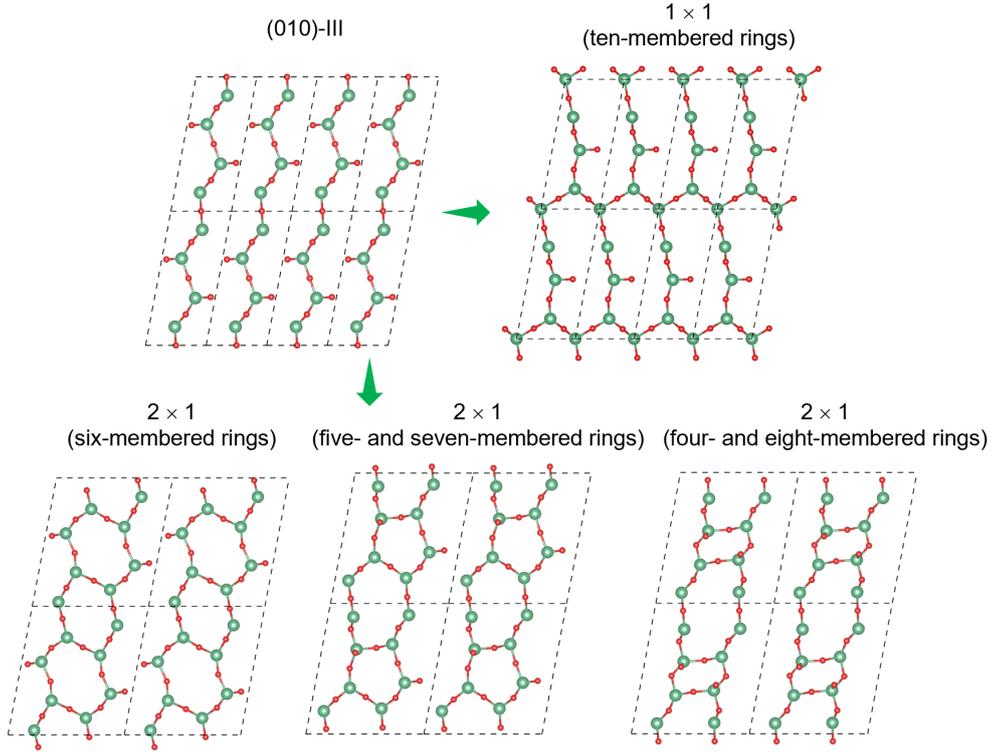

FIG. S5. Possible surface reconstructions for the (010)-III surface.

**Procedures for Searching Possible TLS**

To identify possible TLS that produced by the dangling O atoms, we follow the procedure described below. We first build a large enough supercell, and rotate one dangling O atom away from its original position by 1 Å, without changing the Nb-O bond length. The rotation azimuthal angles are changed from 0° to 360° with a step of 30°. Next, we fix the rotated O atom, as well as the Nb atom that linked to it, and perform a precursive relaxation. After that, we free all the atoms and run a final relaxation. Note that the precursive relaxation of fixing the O and Nb atoms is essential to get the final metastable structure. This is because the rotation of dangling O atom alone cannot form a stable state and it will relax back spontaneously. The precursive relaxation can drive the other atoms to the optimized positions that may help stabilize the new position of the rotated dangling O atom.

## TLS-II on $Nb_2O_5$ and $Ta_2O_5$ Surfaces

The double well potentials of TLS-II on $B$-$Nb_2O_5$ and $B$-$Ta_2O_5$ surfaces are given in Figs. S6, and the corresponding tunnel splittings are listed in Table S2. It can be seen that the tunnel splitting of TLS-II on $B$-$Ta_2O_5$ is almost zero (66 kHz), far below the typical operating frequency (1~10 GHz) of superconducting qubits, and is therefore has negligible effect on the coherence times of Ta qubits. This is mainly caused by the heavier atomic mass of Ta atoms. As shown in Fig. S6(a), the TLS-II primarily involves the hopping of metal atoms (Nb or Ta). According to the Wentzel–Kramers–Brillouin (WKB) theory [24], the tunnel splitting of a symmetric double well potential is very sensitive to the atomic mass $m$ with $E_{TLS} \propto e^{-d\sqrt{2mV/\hbar^2}}$, where $d$ and $V$ are the distance between the two minima and barrier height, respectively. Since the atomic mass of Ta (180.9 amu) is almost two times of Nb (92.9 amu), the tunnel splitting of TLS-II can be significantly different on $B$-$Nb_2O_5$ and $B$-$Ta_2O_5$ surfaces. In fact, by simply changing the atomic mass from Nb to Ta in TLS-II of $B$-$Nb_2O_5$ (Fig. S6(b)), the tunnel splitting dramatically reduced by a factor of 1000, from 0.3 GHz to 0.003 GHz as show in Table S2. Besides, the barrier height of TLS-II on $B$-$Ta_2O_5$ is higher, which further reduces the tunnel splitting to $6.6 \times 10^{-5}$ GHz. Hence, the heavier atomic mass of Ta also plays an important role in the longer coherence time of Ta qubits.

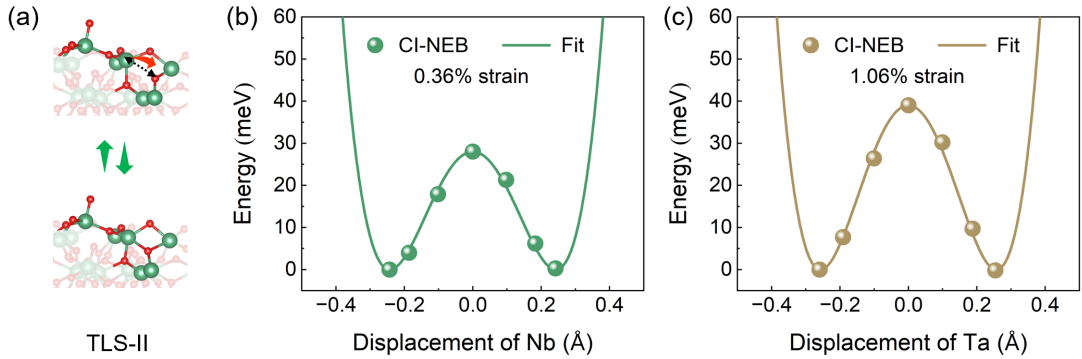

FIG. S6. Double well potential of TLS-II on the $B$-$Nb_2O_5$ and $B$-$Ta_2O_5$ surfaces. (a) Side view of the two local stable states in TLS-II. The metal (Nb or Ta) and O atoms are represented by green and red balls, respectively. The tunneling between the two states mainly involves the hopping of metal atoms. (b) (c) Double well potential of TLS-II at 0.36% (1.06%) strain for $B$-$Nb_2O_5$ ($B$-$Ta_2O_5$).

TABLE S2. Calculated parameters of the double well potential of TLS-II on the $B$-$Nb_2O_5$ and $B$-$Ta_2O_5$ surfaces, and the corresponding tunnel splitting $E_{TLS}$. The third row (Nb → Ta) is the results that simply changing the atomic mass from Nb to Ta in TLS-II of the $B$-$Nb_2O_5$ surface.

| System | Barrier width (Å) | Barrier height (meV) | Atomic mass (amu) | $E_{TLS}$ (GHz) |
| --- | --- | --- | --- | --- |
| $B$-$Nb_2O_5$ | 0.485 | 28 | 92.9 | 0.3 |
| Nb → Ta | 0.485 | 28 | 180.9 | 0.003 |
| $B$-$Ta_2O_5$ | 0.514 | 39 | 180.9 | $6.6 \times 10^{-5}$ |

## Electric Field Tuning of the TLS

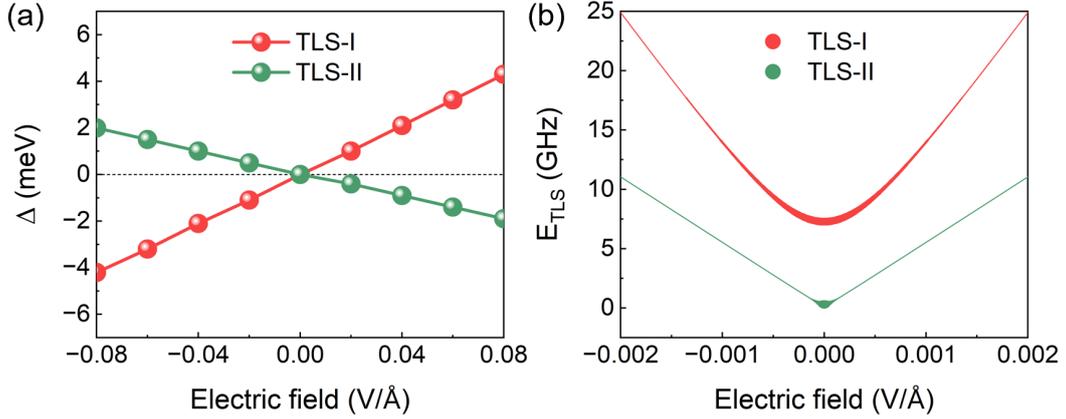

FIG. S7. (a) The asymmetry energy $\Delta$ as a function of external DC electric fields for the two typical TLS. (b) The tunneling splitting of the two TLS tuned by electric fields. The circles size is proportional to the dielectric absorption coefficient $\alpha \propto \mathbf{p'}^2$ [25] where $\mathbf{p'} = \langle \psi_+ | \mathbf{D} | \psi_- \rangle$. Here, $|\psi_+\rangle$ and $|\psi_-\rangle$ are the numerically solved wavefunctions of the TLS, and $\mathbf{D}$ is the electric dipole moment. This is equivalent to $\alpha \propto \Delta_0^2 / E_{TLS}^2$ as demonstrated below.

## Dielectric Absorption Coefficient

In the left and right well basis, the Hamiltonian of an TLS in the perturbing electric field can be expressed as [26,27]

$$H = H_0 + \mathbf{D} \cdot \mathbf{F} = \frac{1}{2} \begin{pmatrix} \Delta + \mathbf{D} \cdot \mathbf{F} & \Delta_0 \\ \Delta_0 & -\Delta - \mathbf{D} \cdot \mathbf{F} \end{pmatrix}, \tag{S6}$$

where **D** is the dipole moment and **F** is the external electric field. While in the TLS basis of

$$|\psi_+\rangle = \sin\left(\frac{\theta}{2}\right)|L\rangle + \cos\left(\frac{\theta}{2}\right)|R\rangle,$$
$$|\psi_-\rangle = \cos\left(\frac{\theta}{2}\right)|L\rangle - \sin\left(\frac{\theta}{2}\right)|R\rangle, \quad (S7)$$

the Hamiltonian is transformed to

$$H = \frac{1}{2}\begin{pmatrix} \sqrt{\Delta^2+\Delta_0^2} & 0 \\ 0 & -\sqrt{\Delta^2+\Delta_0^2} \end{pmatrix} + \frac{1}{2}\frac{1}{\sqrt{\Delta^2+\Delta_0^2}}\begin{pmatrix} \Delta \mathbf{D}\cdot\mathbf{F} & -\Delta_0 \mathbf{D}\cdot\mathbf{F} \\ -\Delta_0 \mathbf{D}\cdot\mathbf{F} & -\Delta \mathbf{D}\cdot\mathbf{F} \end{pmatrix}$$
$$= \frac{1}{2}\begin{pmatrix} E & 0 \\ 0 & -E \end{pmatrix} + \frac{1}{2}\begin{pmatrix} \mathbf{p} & 2\mathbf{p}' \\ 2\mathbf{p}' & -\mathbf{p} \end{pmatrix}\cdot\mathbf{F}, \quad (S8)$$

where

$$\mathbf{p} = \frac{\Delta \mathbf{D}}{\sqrt{\Delta^2+\Delta_0^2}}, \quad \mathbf{p}' = -\frac{1}{2}\frac{\Delta_0 \mathbf{D}}{\sqrt{\Delta^2+\Delta_0^2}}. \quad (S9)$$

It is known that the dielectric absorption coefficient α is proportional to the square of **p**′ [25]. We then obtain

$$\alpha \propto \mathbf{p}'^2 = \frac{1}{4}\frac{\Delta_0^2}{\Delta^2+\Delta_0^2}\mathbf{D}^2 \propto \frac{\Delta_0^2}{E_{\text{TLS}}^2}. \quad (S10)$$